\documentclass[runningheads]{llncs}
\usepackage[T1]{fontenc}
\usepackage{graphicx}

\usepackage{amsmath,amsfonts,bm}

\def\Figref#1{Figure~\ref{#1}}

\def\Secref#1{Section~\ref{#1}}

\def\eqref#1{equation~\ref{#1}}
\def\Eqref#1{Equation~\ref{#1}}

\def\1{\bm{1}}

\def\rvb{{\mathbf{b}}}

\def\rmW{{\mathbf{W}}}

\def\vb{{\bm{b}}}

\def\vf{{\bm{f}}}

\def\vs{{\bm{s}}}

\def\vv{{\bm{v}}}

\def\vx{{\bm{x}}}
\def\vy{{\bm{y}}}

\def\mO{{\bm{O}}}

\def\mR{{\bm{R}}}

\def\mV{{\bm{V}}}
\def\mW{{\bm{W}}}

\DeclareMathAlphabet{\mathsfit}{\encodingdefault}{\sfdefault}{m}{sl}
\SetMathAlphabet{\mathsfit}{bold}{\encodingdefault}{\sfdefault}{bx}{n}

\def\gE{{\mathcal{E}}}

\def\gG{{\mathcal{G}}}

\def\gV{{\mathcal{V}}}

\def\sR{{\mathbb{R}}}

\newcommand{\R}{\mathbb{R}}

\usepackage{hyperref}
\usepackage{url}
\usepackage{subfigure}
\usepackage{booktabs} 
\usepackage{url}
\usepackage{multirow}
\usepackage{tabularx}

\begin{document}
\title{Orientation-Aware Graph Neural Networks for Protein Structure Representation Learning}
%
%
\author{
Jiahan Li$^*$\inst{1,2} \and
Shitong Luo$^*$\inst{3} \and
Congyue Deng$^*$\inst{4} \and
Chaoran Cheng$^*$\inst{5} \and
Jiaqi Guan\inst{5} \and
Leonidas Guibas\inst{4} \and
Jian Peng\inst{6} \and
Jianzhu Ma\inst{2,6}
}
\authorrunning{Li et al.}
%
\institute{Peking University \and
Tsinghua University \and
Massachusetts Institute of Technology \and
Stanford University \and
University of Illinois Urbana-Champaign \and
Helixon Ltd. \\
\email{ced3ljhypc@gmail.com}
}
\maketitle              

\begin{abstract}
By folding into particular 3D structures, proteins play a key role in living beings. To learn meaningful representation from a protein structure for downstream tasks, not only the global backbone topology but the local fine-grained orientational relations between amino acids should also be considered. In this work, we propose the Orientation-Aware Graph Neural Networks (OAGNNs) to better sense the geometric characteristics in protein structure (e.g. inner-residue torsion angles, inter-residue orientations). Extending a single weight from a scalar to a 3D vector, we construct a rich set of geometric-meaningful operations to process both the classical and SO(3) representations of a given structure. To plug our designed perceptron unit into existing Graph Neural Networks, we further introduce an equivariant message passing paradigm, showing superior versatility in maintaining SO(3)-equivariance at the global scale. Experiments have shown that our OAGNNs have a remarkable ability to sense geometric orientational features compared to classical networks. OAGNNs have also achieved state-of-the-art performance on various computational biology applications related to protein 3D structures. The code is available at \url{https://github.com/Ced3-han/OAGNN/tree/main}.

\end{abstract}

\newpage

\section{Introduction}

Built from a sequence of amino-acid residues, a protein performs its biological functions by folding to a particular conformation in 3D space. Therefore, utilizing such 3D structures accurately is the key to downstream analysis. While we have witnessed remarkable progress in structure predictions \cite{rohl2004protein,kallberg2012template,baek2021accurate,jumper2021highly}, another thread of tasks with protein 3D structures as input, starts to draw a great interest, such as function prediction \cite{hermosilla2020intrinsic,gligorijevic2021structure}, decoy ranking \cite{lundstrom2001pcons,kwon2021assessment,wang2021protein}, protein docking \cite{duhovny2002efficient,shulman2004recognition,gainza2020deciphering,sverrisson2021fast}.

\begin{figure}[t]
    \centering
    \includegraphics[width=0.65\linewidth]{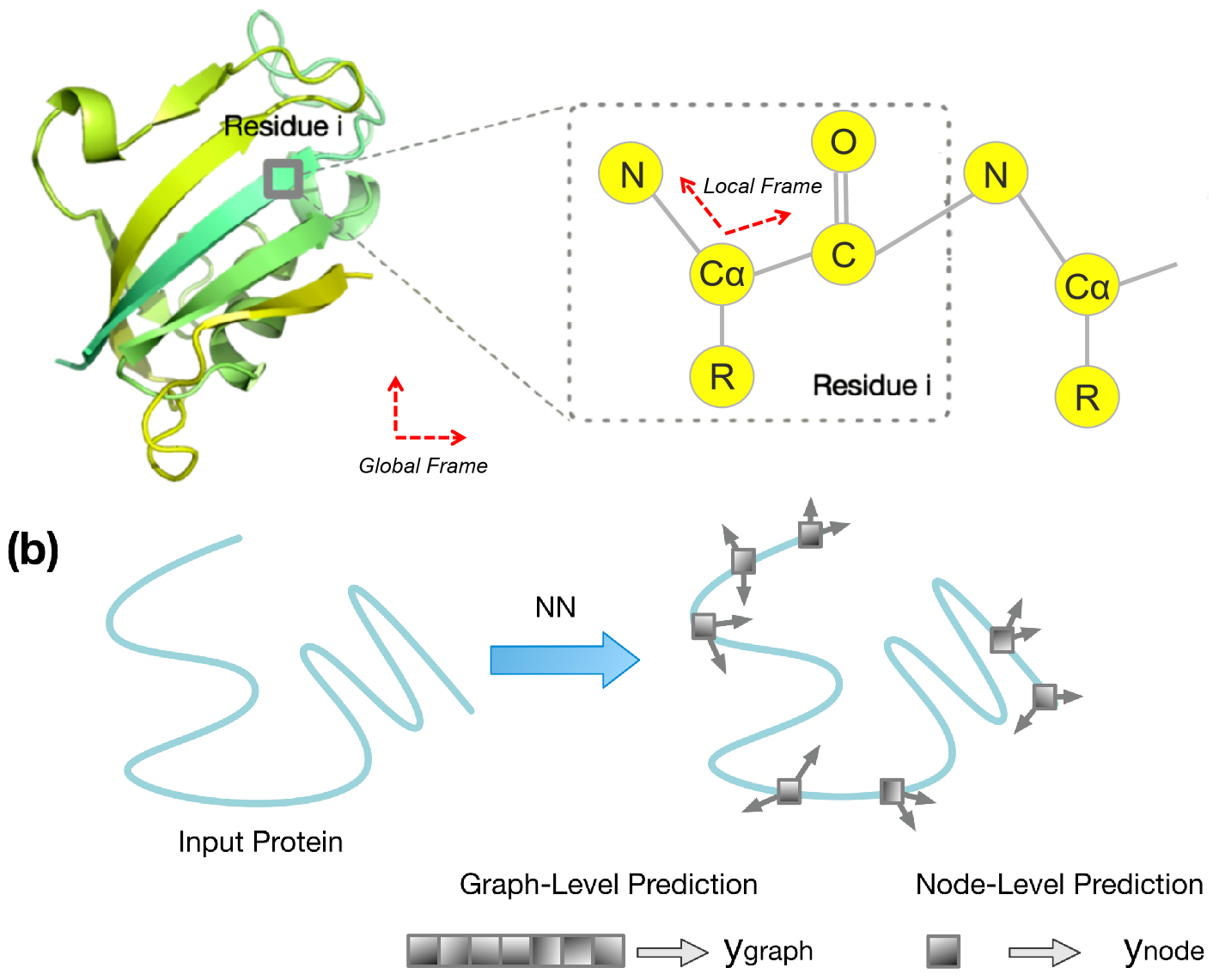}
    \vspace{-1.5em}
    \caption{
    Overview (a) Each amino acid has its own \textbf{rigid backbone} with four heavy atoms, representing a local frame. (b) Tasks associated with the protein 3D structure. \textbf{Graph-Level} tasks consider the whole protein structures, and \textbf{Node-Level} tasks operate on specific residues.
    }
    \vspace{-1.5em}
    \label{fig:model_overview}
\end{figure}

Most existing works in modeling protein structures directly borrow models designed for other applications, including 3D-CNNs \cite{ji20123d}, Transformers \cite{vaswani2017attention}) and GNNs \cite{kipf2016semi}. However, these models have overlooked the subtleties in the fine-grained geometries, which are much more essential in protein structures. For instance, given an amino acid in the protein structure, as shown in \Figref{fig:model_overview}, the locations of four backbone atoms determine a local frame and different residues interact with each other through performing specific orientations between their frames, either of which have significant impacts on the protein structure and its function \cite{nelson2008lehninger}. 

Recent attempts in building geometry-aware neural networks mainly focus on baking \emph{3D rigid transformations} into network operations, leading to the area of $\mathrm{SO}(3)$-invariant and equivariant networks. One representative work is the Vector Neuron Network (VNN) \cite{deng2021vector}, which achieves $\mathrm{SO}(3)$-equivariance on point clouds by generalizing scalar neurons to 3D vectors. Another work is the GVP-GNN \cite{jing2021equivariant} that similarly vectorizes hidden neurons in GNN and demonstrates better accuracy on protein tasks. However, they can only adopt \emph{linear combinations} of input vectors, significantly limiting their modeling capability. A simple example is that given two input vector features $\vv_1$ and $\vv_2$, the outputs $w_1\vv_1+w_2\vv_2$ through one linear layer are constrained in the 2D plane spanned by $\vv_1,\vv_2$ even after applying their scalar-product non-linearities. That is, VNN-based models are limited in processing orientational features, which have been proven crucial for proteins to perform their functions and interact with other partners (\textit{e.g.} inner-residue torsion angles, inter-residue orientations) \cite{nelson2008lehninger,voet2010biochemistry,xu2006fast,alford2017rosetta}.

To achieve more sensitive geometric orientation awareness, we propose a \emph{Directed Weight ($\vec{\rmW}$) perceptrons} by extending not only the hidden neurons but also the weights from scalars to 3D vectors, naturally saturating the entire network with 3D structure information in the Euclidean space. Directed weights support a set of geometric-meaningful operations on both the vector neurons (vector-list features) and the classical (scalar-list) latent features and perform flexible non-linear integration of the hybrid scalar-vector features. As protein structures are naturally attributed proximity graphs, we introduce a new \emph{Equivariant Message Passing Paradigm} on protein graphs to connect the $\vec{\rmW}$-perceptrons with the graph learning model by using rigid backbone transformations for each residue, which provides a versatile framework for bringing the biological suitability and flexibility of the GNN architecture. 

To summarize, our key contributions include: (1) We propose a new network unit based on the \emph{Directed Weights} for capturing fine-grained geometric relations, especially for proteins; (2) We construct an \emph{Equivariant Message Passing} paradigm on protein graphs; (3) Our overall framework, the \emph{Orientation-Aware Graph Neural Networks}, is versatile in terms of compatibility with existing deep graph learning models.
\section{Directed Weight Perceptron}
\label{dwp}

A protein is modeled as a KNN-graph $\gG = (\gV,\gE)$ where each node $u\in\gV$ corresponds to one residue, characterized by a scalar-vector tuple $ h_u = (\vs_u,\mV_u)$ with $\vs_u \in \mathbb{R}^{C}, \mV_u \in \mathbb{R}^{C \times 3}$, and the edges are constructed spatially by querying its $k$-nearest neighbours in 3D. The edge features $\mathcal{E}=\{e_{ij}\}_{i\neq j}$ representing edge connected node $i$ and $j$, are multi-channel scalar-vector tuples. For simplicity, we set the channel numbers for scalar and vector features as the same, but they can be different in practice. Details can be found in the Appendix \Secref{pf}.


\begin{figure*}[!htbp]
    \centering
    \includegraphics[width=\linewidth]{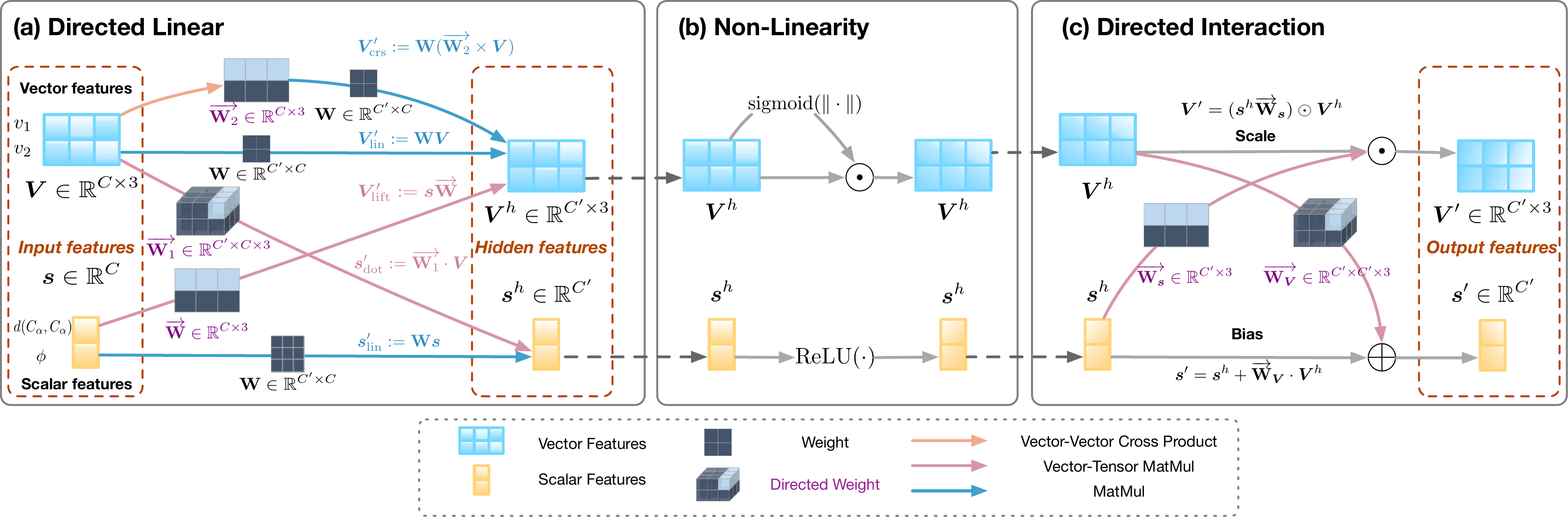}
    \vspace{-2.0em}
    \caption{
    Model Details. A 1-layer DWP consists of three following modules.
    (a) \textbf{Directed Linear Module} applies multiple geometric operations to update scalar and vector features in four different ways with directed weights. 
    (b) \textbf{Non-Linearity Module} employs ReLU and sigmoid functions for scalar and vector features.
    (c) \textbf{Directed Interaction Module} updates scalar and vector features by using one another as updating parameters after non-linearity.
    }
    \vspace{-2.0em}
    \label{fig:model_detail}
\end{figure*}

\subsection{Directed Weights}
\label{dwp:weights}

Classical neural networks only consider scalar-list features of the form $\vs \in \sR^C$, with each layer transforming them with a weight matrix $\rmW \in \sR^{C'\times C}$ and a bias term $\vb\in\sR^{C'}$:
\begin{equation}
    \vs' = \rmW \vs + \rvb
\end{equation}
Although it has been proven to be a universal approximator, such a layer has intrinsically no geometric interpretation in 3D space, and even the simplest 3D rigid transformation on the input can result in unpredictable changes in network outputs. Recent attempts have lifted neuron features from scalar-lists to vector-lists $\mV\in\sR^{C\times3}$ with a linear operation mapping to $\mV'\in\sR^{C'\times3}$ \cite{deng2021vector,jing2020learning}:
\begin{equation}
    \mV' = \rmW\mV
\end{equation}
Under this construction, $\mathrm{SO}(3)$-actions in the latent space are simply matrix multiplications, establishing a clear 3D Euclidean structure. However, to maintain input-output consistency under rigid transformations (e.g., rotation and translation), their operations are still limited to classical linear weighted combinations. To define \emph{operations} that are more adaptive to the geometrically meaningful features, we introduce the \emph{Directed Weights}, that we can define any tensor with the last dimension of $3$ as a directed weight matrix, for example, $\overrightarrow{\rmW} \in \sR^{C'\times C\times3}$, seen as a stack of geometric meaningful vectors lying in the 3D Euclidean Space, and can be acted by $\mathrm{SE}(3)$ group from the right.



\subsection{$\overrightarrow{\rmW}$-Operators}
\label{dwp:operators}

With weights and features both equipped with 3D vector representations, we can design a set of \emph{geometric operators} $\square$ (e.g. $\cdot$ or $\times$) with learnable $\overrightarrow{\rmW}$ as parameters in neural networks, which can operate on both scalar-list features $\vs \in \sR^C$ and vector-list features $\mV \in \sR^{C\times3}$
\begin{equation}
    \overrightarrow{\rmW}\square\vs, \quad \overrightarrow{\rmW}\square \mV
\end{equation}

\textbf{Geometric vector operations.} 
Let $\rmW \in  \sR^{C'\times C}$ be a conventional scalar weight matrix, $\overrightarrow{\rmW}_1 \in \sR^{C'\times C\times 3}$ and $\overrightarrow{\rmW}_2 \in \sR^{C\times 3}$ be directed weight tensors. Beyond linear combinations, We can define two operations that leverage geometric information:
\begin{alignat}{3}
    \vs_{\text{dot}}'(\mV; \overrightarrow{\rmW}_1) &= \overrightarrow{\rmW}_1 \cdot \mV
    &&\quad\in \mathbb{R}^{C'} \label{eqn:vs}
    \\
    \mV_{\text{crs}}'(\mV; \rmW, \overrightarrow{\rmW}_2) &= \rmW (\overrightarrow{\rmW}_2 \times \mV)
    &&\quad\in \R^{C' \times 3} \label{eqn:vvc}
\end{alignat}
Here $\vs_{\text{dot}}'$ transforms $C$ vector features to $C'$ scalars using dot-product with directed weights, detailedly, $\vs'_{i} = \sum_{j,k}\vec{\mW}_1^{i,j,k}\mV^{j,k}$, which explicitly measures angles between vectors, and this operator brings network the ability to accurately sense the orientational features. In $\mV_{\text{crs}}'$, a vector crosses a directed weight before being projected onto the hidden space. While the output of plain linear combinations between two vectors $\vv_1$ and $\vv_2$ can only lie in the plane $w_1\vv_1+w_2\vv_2$, cross-product in \Eqref{eqn:vvc} creates a new direction outside the plane, which is crucial in 3D modelling. 


\textbf{Scalar lifting.}
In addition, a directed weight $\overrightarrow{\rmW} \in \sR^{C\times 3}$ can lift scalars to vectors by adopting the following operation, the $ij$ th entry of $s\overrightarrow{W}$ is $s_i\overrightarrow{W}_{ij}$
\begin{alignat}{3}
    \mV_{\text{lift}}'(\vs; \overrightarrow{\rmW}) &= \vs\overrightarrow{\rmW}
    &&\quad\in \mathbb{R}^{C \times 3}
    \label{eqn:sv}
\end{alignat}
This maps each scalar to a particular vector, enabling inverse transformations or information bottlenecks from $\mathbb{R}$ to $\mathbb{R}^3$. An intuition example is that, we can map a scalar representing the distance between two residues to a vector pointing from one to the other, leading to more biological meaningful representations.

\textbf{Linear combinations.}
In the end, we keep the linear combination operations with scalar weights for both scalar and vector features:
\begin{alignat}{3}
    \vs_{\text{lin}}'(\vs; \rmW) &= \rmW\vs
    &&\quad\in \R^{C'} \label{eqn:ss}
    \\
    \mV_{\text{lin}}'(\mV; \rmW) &= \rmW\mV
    &&\quad\in \R^{C' \times 3} \label{eqn:vvs}
\end{alignat}
While these operators enables a more flexible network design, equivariance to rigid transformations is broken if considering more complex functions beyond linear combinations. We will introduce a globally equivariant paradigm to tackle this issue in Section \ref{dwg}.

\subsection{$\overrightarrow{\rmW}$-Perceptron Unit}
\label{dwp:percep_unit}

Now we combine all the $\overrightarrow{\rmW}$-operators together, assembling them into what we call Directed Weight Perceptrons ($\overrightarrow{\rmW}$-perceptrons). A $\overrightarrow{\rmW}$-perceptron unit is a function mapping from $u = (\vs,\mV)$ to another scalar-vector tuple $u' = (\vs',\mV')$, which can be stacked as network layers to form multi-layer perceptrons. 

A single unit comprises three modules in a sequential way: the \textbf{Directed Linear} module, the \textbf{Non-Linearity} module, and the \textbf{Directed Interaction} module (\Figref{fig:model_detail}).

\textbf{Directed linear module.}
The output hidden representations derived from the set of operations introduced previously are concatenated and projected into another hidden space of scalars and vectors separately:
\begin{alignat}{3}
    \vs^h &= \rmW_\vs [\vs_{\text{dot}}',\, \vs_{\text{lin}}']
    && \quad\in \R^{C'} \\
    \mV^h &= \rmW_\mV [\mV_{\text{crs}}',\, \mV_{\text{lift}}',\, \mV_{\text{lin}}']
    && \quad\in \R^{C'\times 3}
\end{alignat}

Here $\rmW_\vs \in \mathbb{R}^{C'\times(C'+C')}$ and $\rmW_\mV \in \mathbb{R}^{C'\times(C'+C +C')}$ are scalar weight matrices. In other word, the five separate updating functions allow transformation from scalar to scalar (\Eqref{eqn:ss}), scalar to vector (\Eqref{eqn:sv}), vector to scalar (\Eqref{eqn:vs}), and vector to vector (\Eqref{eqn:vvs}, \Eqref{eqn:vvc}),  boosting the model’s capability to reason in 3D space.

\textbf{Non-linearity module.}
We then apply the non-linearity module to the hidden representation $(\vs^h,\mV^h)$. Specifically, we apply standard ReLU non-linearity \cite{nair2010rectified} to the scalar components. For the vector representations, following \cite{weiler20183d,jing2020learning,schutt2021equivariant}, we compute a sigmoid activation on the L2-norm of each vector and multiply it back to the vector entries accordingly:
\begin{align}
    \vs^h \gets \mathrm{ReLU}(\vs^h),\quad
    v_{ij}^h \gets v_{ij}^h\cdot\mathrm{sigmoid}(\|\vv_i^h\|_2)
\end{align}
where $\vv_i^h \in \sR^3$ are the vector columns in $\mV^h$ and $v_{ij}^h$ are their entries.

\textbf{Directed interaction module.}
Finally we introduce the Directed Interaction module, integrating the hidden features $\vs^h$ and $\mV^h$ into the output tuple $(\vs',\mV')$
\begin{alignat}{3}
    \vs' &= \vs^h + \overrightarrow{\rmW}_\mV \cdot \mV^h   &&\quad\in \R^{C'} \label{eqn:interact:s}
    \\
    \mV' &= (s^h \overrightarrow{\rmW}_\vs) \odot \mV^h   &&\quad\in \R^{C'\times 3} \label{eqn:interact:v}
\end{alignat}
Here $\overrightarrow{\rmW}_\mV, \overrightarrow{\rmW}_\vs$ are directed weight matrices with sizes $C' \times C'\times 3$ and $C' \times 3$, respectively, $\odot$ denotes element-wise multiplication for two matrices.
This module establishes a connection between the scalar and vector feature components, facilitating feature blending. Specifically, \Eqref{eqn:interact:s} dynamically determines how much the output should rely on scalar and vector representations, and \Eqref{eqn:interact:v} weights a list of vectors using the scalar features as attention scores. 

\section{Orientation-Aware Graph Neural Networks}
\label{dwg}

To achieve $\mathrm{SO}(3)$-equivariance without the loss of modeling capacity, we introduce an \emph{equivariant message passing paradigm} on protein graphs, to easily plug our versatile $\overrightarrow{\rmW}$-Perceptron into any existing graph learning framework, which makes network architectures free from equivariant constraints (\Secref{emp}). 
The integrated models, called \emph{Orientation-Aware Graph Neural Networks}, can not only accurately model the essential orientational features but also maintain the rotation equivariance efficiently. We also design multiple variants in analogy to other graph neural networks \Secref{variants}.

\subsection{Equivariant Message Passing on Proteins}
\label{emp}

A function $\vf:\R^3 \to \R^3$ is $\mathrm{SO}(3)$-equivariant (rotation-equivariant) if any rotation matrix $\mR \in \R^{3\times 3}$ applied to the input vector $\vx$ leads to the same transformation on the output $\vf(\vx)$:
\begin{equation}
    \mR\vf(\vx) = \vf(\mR\vx)
\end{equation}
Such equivariant property can be achieved on a protein graph with local orientations, which is naturally defined from its biological structure \cite{ingraham2019generative}.
Specifically, each amino acid node $u_i\in\gV$ has four backbone heavy atom ($C_\alpha^u,C^u,N^u,O^u$), defining a local frame $\mO_u$ as:
\begin{align}
    \vx_u &= N^u-C_\alpha^u \in \mathbb{R}^3, \quad
    \vy_u = C^u-C_\alpha^u \in \mathbb{R}^3 \\
    \mO_u &=\left[ \frac{\vx_u}{\|\vx_u\|_2},\frac{\vy_u}{\|\vy_u\|_2},\frac{\vx_u}{\|\vx_u\|_2} \times \frac{\vy_u}{\|\vy_u\|_2}\right]^\top \in \mathbb{R}^{3\times 3}
    \label{frame}
\end{align}
The local frame $\mO_u$ is a rotation matrix that maps a 3D vector from the local to the global coordinate system.
An equivariant message passing paradigm then emerges through transforming node features back and forth between adjacent local frames.
Formally, give an amino acid $u$ with hidden representation $h=(\vs,\mV)$, let $f_l$ and $f_g$ be transformations on the vector feature $\mV$ from and to the global coordinate system:
\begin{align}
    f_l(h,\mO_u) := (\vs,\mV \mO_u^\top), \quad
    f_g(h,\mO_u) := (\vs,\mV \mO_u)
\end{align}
The message passing update for node $u$ from layer $l$ to layer $l+1$ is performed in the following steps, the $[\quad]$ notations indicates  the concatenation of different hidden representations along the  channel-wise dimension:
\begin{enumerate}
    \item Transform the neighbourhood vector representations of $u$ into its local coordinate system $\mO_u$.
    \item For each adjacent node $v$, compute the message $m^{l+1}_u$ on edge $(u,v)\in\mathcal{E}$ with an multi-layer $\overrightarrow{\rmW}$-perceptron $\mathcal{F}$ to the aggregated node and edge feature $[h_u^l, h_v^l, e_{uv}^l]$ (\Eqref{eqn:message}).
    \item Update node feature $h_u^l$ with another multi-layer $\overrightarrow{\rmW}$-perceptron $\mathcal{H}$, and transform it back to the global coordinate system (\Eqref{eqn:update}).
\end{enumerate}
This paradigm achieves $\mathrm{SO}(3)$-equivariance in a very general sense with no constraints on $\mathcal{F},\mathcal{H}$. The formal proof of equivariance is presented in the Appendix \Secref{pr}. 
\begin{align}
    m_{u}^{l+1}&=\sum_{v\in\mathcal{N}_u} \mathcal{F}(f_l([h_u^l, h_v^l, e_{uv}^l], \mO_u))\label{eqn:message}
    \\
    h_u^{l+1}&=f_g(\mathcal{H}(h_u^l, m_u^{l+1}),\mO_u)
    \label{eqn:update}
\end{align}

\subsection{Variants of Orientation-Aware Graph Neural Networks}
\label{variants}
By integrating the $\overrightarrow{\rmW}$-Perceptrons with equivariant message passing, we propose multiple variants of entire \emph{Orientation-Aware Graph Neural Networks} to boost the design space for different tasks.

\textbf{OA-GCN}. This is the one described and implemented in the previous subsection with \Eqref{eqn:message} and \Eqref{eqn:update}. 

\textbf{OA-GIN}. We adopt graph isomorphism operator \cite{xu2018powerful} with learnable weighing parameter $\varepsilon$ for tunable skip connections,
\begin{align}
    m_u^{l+1}&=(1+\varepsilon)f_l(h_u^l,\mO_u)+\sum_{v\in\mathcal{N}_u} \mathcal{F}(f_l(h_v^l,\mO_u))
\end{align}
\textbf{OA-GAT}. In comparison to GCN and GIN, it is not trivial to incorporate with the Graph Attention Network (GAT) \cite{velivckovic2018graph}, as residues interact with each other unevenly in proteins. In particular, we define separate attentions for the scalar and vector representations
\begin{align}
    (\vs_u^l,\mV_u^l)&= f_l(h_u^l,\mO_u)\\
    \vs'_u&=\sum_{v\in\mathcal{N}_u\cup\{u\}}\alpha^s_v \vs_v^l
    \quad\mV'_u=\sum_{v\in\mathcal{N}_u\cup\{u\}}\alpha^v_v \mV_v^l \\
    h_u^{l+1}&=f_g(\mathcal{H}((\vs',\mV')),\mO_u)
\end{align}
The attention scores $\alpha^s,\alpha^v$ are the softmax values over the inner products of all neighboring source-target pairs defined as the follows:
\begin{align}
    \alpha^s_v=\frac{\exp{(\langle \vs_u,\vs_v\rangle)}}{\sum_{w\in\mathcal{N}_u\cup\{u\}}\exp{(\langle \vs_u,\vs_w\rangle)}},
    \quad
    \alpha^v_v=\frac{\exp{(\mathrm{tr}(\mV_u^\top \mV_v)})}{\sum_{w\in\mathcal{N}_u\cup\{u\}}\exp{(\mathrm{tr}(\mV_u^\top \mV_v))}} \label{eqn:attentionV}
\end{align}
The product for scalars and vectors are standard inner product and Frobenius product, respectively. 
%

\section{Experiments}
\label{dexp}

To demonstrate the basic rationale in perceiving angular features of our $\overrightarrow{\rmW}$-Perceptron, we design a synthetic task (\Secref{syn}). We conduct experiments including node-level Residue Identification (\textbf{RES}) (\Secref{resi}), Computational Protein Design (\textbf{CPD}) (\Secref{scpd}), and graph-level Model Quality Assessment (\textbf{MQA}) and Protein Classification(Appendix \Secref{pfc}). We also conduct ablation studies of different components based on OA-GCN; we replace directed weights with the scalar-weight linear layer (\textbf{No DW}), or remove the Interaction Module (\textbf{No Int}), or break the equivariance by canceling coordinate transformations during message passing (\textbf{No Equi}). Details are presented in Appendix \Secref{md}.



 \subsection{Synthetic Task}
\label{syn}


\begin{figure}[t]
    \centering
    \includegraphics[width=0.9\linewidth,height=0.4\linewidth]{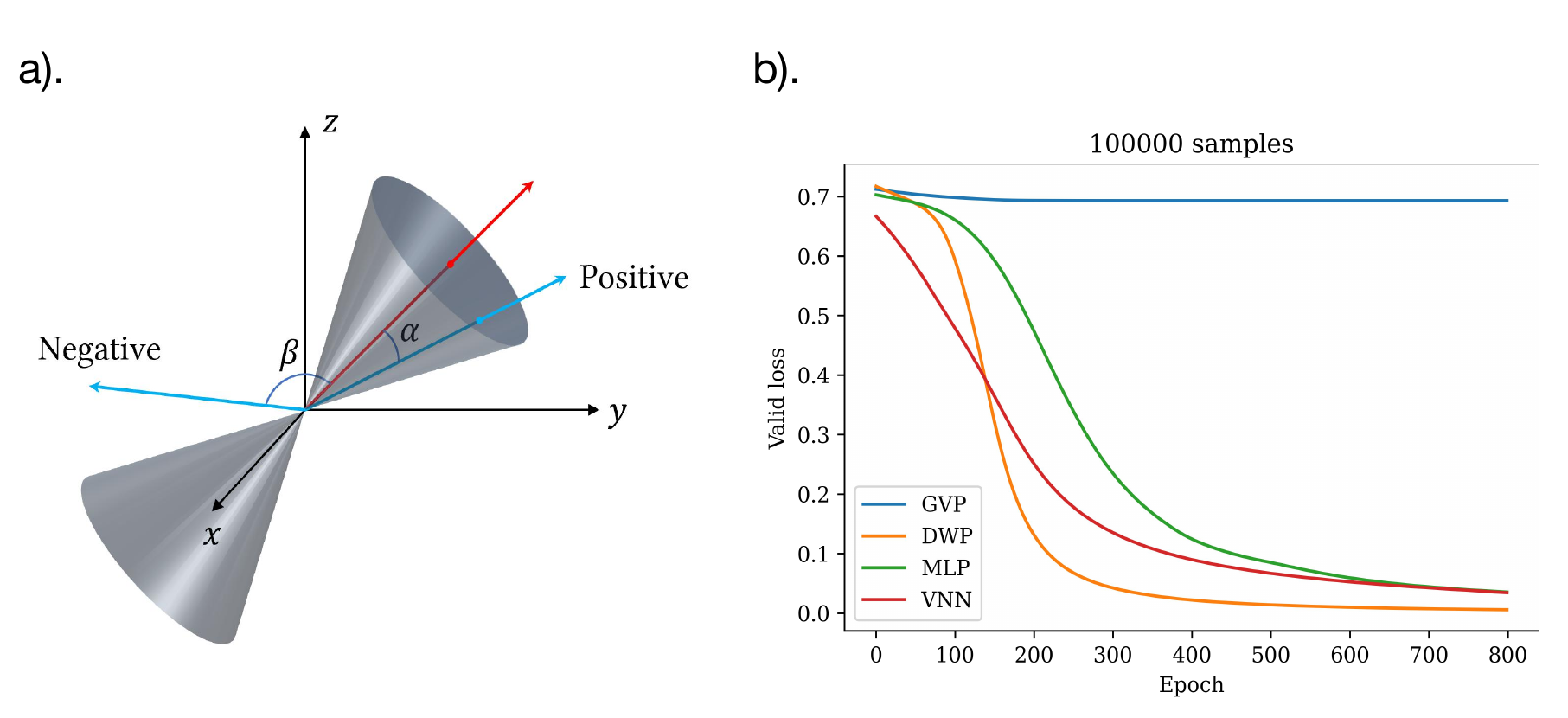}
    \vspace{-1.8em}
    \caption{a). \textbf{The synthetic study}. The vector in red is the anchor vector. Other vectors falling into the cone are positive, whereas outside points are negative. b). The validation Binary Cross Entropy loss of compared methods.}
    \label{fig:toy}
    \vspace{-0.7em}
\end{figure}


We test the performance of different perceptron designs on a synthetic angular binary classification task. We sample a random unit anchor vector $\vv_m \in \mathbf{R}^3$. The other vector $\vv_i$ is labeled positive if its angle between the anchor $a_{mi} \leq \pi/4$, and negative otherwise. We generate $100,000$ random vectors with $50\%$ positive and $50\%$ negative. For a fair comparison, we adjust the number of parameters and hyperparameters of different models to approximately the same.

\textbf{Results.} As shown in \Figref{fig:toy}, in general, our $\overrightarrow{\rmW}$-Perceptron (\textbf{DWP}) converges faster than other models and achieves the lowest validation loss. We also notice that the Geometric Vector Perceptron (\textbf{GVP}) cannot converge given a sufficient number of epochs due to the poor capability of perceiving angulars by only using linear combinations of vector. Compared to GVP, Vector Neural Network (\textbf{VNN}) performs better based on its non-linear ReLU vector operation. Simple Multi-Layer Perceptron (\textbf{MLP}) also captures useful information according to its universal approximation ability. The superior performance demonstrates that $\overrightarrow{\rmW}$-Perceptron has a better geometric reasoning capability.

\subsection{Residue Identification}
\label{resi}

\begin{table}[t]
\begin{minipage}[t]{.67\textwidth}
    \centering
    \caption{Results of different RES methods on ATOM3D.}\label{table:resr}
    \vspace{-0.2em}
    \resizebox{\linewidth}{!}{
    \begin{tabular}{@{}l|ccccccccc@{}}
    \toprule
    Model & 3D-CNN & GCN & GIN & GAT & GVP-GNN & OA-GCN & OA-GIN & OA-GAT \\ \midrule
    Acc \% & 45.1 & 8.2 & 9.1 & 12.4 & 48.2 & 50.2 & 50.8 & 49.2 \\ \bottomrule
    \end{tabular}
    }
\end{minipage}
\hfill
\begin{minipage}[t]{.31\textwidth}
    \centering
    \caption{Ablation studies.}\label{table:resr_abl}
    \vspace{-0.25em}
    \resizebox{\linewidth}{!}{
    \begin{tabular}{@{}l|cccc@{}}
    \toprule
    Model & No DW & No Int & No Equi &  \\ \midrule
    Acc \% & 47.3 & 47.7 & 33.0 & \\ \bottomrule
    \end{tabular}
    }
\end{minipage}
\vspace{-2em}
\end{table}

Residue Identification (RES) aims to predict the identity of a particular masked amino acid based on its local surrounding structure \cite{torng20173d}. We download $10^6$ substructures from the ATOM3D project \cite{townshend2020atom3d}. The entire dataset is split into training, validation, and testing datasets with the ratio of $80\%$, $10\%$, and $10\%$. There are no similar structures between test and non-test datasets \cite{orengo1997cath}


\textbf{Metrics and Baselines} We use classification accuracy to evaluate model performance. The \textbf{3D-CNN} encodes the positions of the atoms in a voxelized 3D volume. We also compare GNN variants such as \textbf{GCN}, \textbf{GIN} and \textbf{GAT} with our \textbf{OA-GCN}, \textbf{OA-GIN} and \textbf{OA-GAT} models. \textbf{GVP-GNN} is also included.

\textbf{Results} As shown in Table \ref{table:resr}, \textbf{GCN}, \textbf{GIN} and \textbf{GAT} almost cannot accurately identify the type of amino acids, suggesting the lack of ability of conventional GNN models to capture 3D geometric information. \textbf{3D-CNN} performs better due to the power of the voxelization technique. Our models outperform other models, suggesting OA-GNNs could help represent protein substructure geometries better. Here, equivariance message passing is the most essential component, indicating the necessity of rotation equivariance in perceiving local protein structures.

\subsection{Computational Protein Design}
\label{scpd}

Computational Protein Design (CPD) predicts the native protein sequence of a given backbone structure. Specifically, we focus on two databases: CATH 4.2 \cite{ingraham2019generative} organizes proteins in a hierarchical structure \cite{orengo1997cath}. Following prior works, $18,024$ protein chains are in the training set, $608$ in the validation set, and $1,120$ in the test. TS50 dataset \cite{li2014direct} is a relatively old benchmark of only $50$ protein structures widely used in biology communities. \footnote{As there is no canonical training set for TS50, we follow prior methods to use protein sequences with less than 30$\%$ similarity with the TS50 test set from the CATH training set for training \cite{jing2020learning,qi2020densecpd}}.

\textbf{Metrics.} Native Sequence Recovery Rate compares the predicted sequence with the ground truth sequence and calculates the proportion of the correctly recovered residues \cite{li2014direct}. The perplexity score \cite{jelinek1977perplexity} evaluates whether a model could assign a high likelihood to the test sequences \cite{ingraham2019generative}.


\vspace{-1.5em}

\begin{table}[!htbp]
\begin{minipage}[t]{0.52\linewidth}
    \centering
    \caption{Results of different CPD methods on the CATH 4.2.}\label{table:cath}
    \vspace{-0.6em}
    \resizebox{!}{2.4cm}{
    \begin{tabular}{lccccccc}
    \toprule 
    \multirow{2}{*}{Model} & \multicolumn{3}{c}{Perplexity $\downarrow$} & & \multicolumn{3}{c}{Recovery \% $\uparrow$} \\
    \cmidrule(lr){2-4} \cmidrule(lr){6-8}
    & Short & Single & All & & Short & Single & All \\
    \midrule 
    St-Transformer  & 8.54 & 9.03 & 6.85 & & 28.3 & 27.6 & 36.4 \\
    St-GCN  & 8.31 & 8.88 & 6.55 & & 28.4 & 28.1 & 37.3 \\
    St-GIN  & 8.03 & 8.52 & 6.15 & & 27.7 & 28.4 & 38.1 \\
    St-GAT  & 10.86 & 10.67 & 9.89 & & 26.2 & 26.8 & 35.2 \\
    GVP-GNN & 7.10 & 7.44 & 5.29 & & 32.1 & 32.0 & 40.2 \\
    \midrule
    OA-GCN  & \textbf{5.42} & 5.69 & 3.94 & & \textbf{39.6} & 38.5 & 47.5 \\
    OA-GIN  & 5.84 & \textbf{5.39} & \textbf{3.85} & & 38.8 & \textbf{40.1} & \textbf{47.8}\\
    OA-GAT  & 5.92 & 5.53 & 4.13 & & 37.5 & 39.3 & 46.7 \\
    \midrule 
    No DW  & 6.52 & 6.79 & 5.28 & & 36.7 & 35.0 & 42.9 \\
    No Int     & 6.45 & 6.36 & 4.87 & & 37.2 & 36.3 & 44.9 \\
    No Equi     & 6.21 & 6.04 & 4.64 & & 36.9 & 37.1 & 45.8 \\
    \bottomrule 
    \end{tabular}
    }
    \end{minipage}
\hfill
\begin{minipage}[t]{0.45\linewidth}
    \centering
    \caption{Results of MQA models on CASP 11 stage 2.}\label{table:casp}
\vspace{-0.75em}
    \resizebox{!}{2.4cm}{
    \begin{tabular}{lccccccc}
    \toprule 
    \multirow{2}{*}{Model} & \multicolumn{3}{c}{Average $\uparrow$} & & \multicolumn{3}{c}{Global $\uparrow$} \\
    \cmidrule(lr){2-4} \cmidrule(lr){6-8}
    & $r$ & $\rho$ & $\tau$ & & $r$ & $\rho$ & $\tau$ \\
    \midrule 
    VoroMQA & 0.42 & 0.41 & 0.29 & & 0.65  & 0.69 & 0.51 \\
    RWplus  & 0.17 & 0.19 & 0.13 & & 0.06  & 0.03 & 0.01 \\
    SBROD   & 0.43 & 0.41 & 0.29 & & 0.55  & 0.57 & 0.39 \\
    Proq3D  & 0.44 & \textbf{0.43} & 0.30 & & 0.77 & \textbf{0.80} & 0.59 \\
    3DCNN   & 0.49 & 0.39 & 0.27 & & 0.64  & 0.67 & 0.48 \\
    Ornate  & 0.39 & 0.37 & 0.26 & & 0.63  & 0.67 & 0.48 \\
    DimeNet  & 0.30 & 0.35 & 0.28 & & 0.61  & 0.62 & 0.43 \\
    GraphQA  & 0.48 & 0.40 & 0.42 & & 0.75  & 0.72 & 0.74 \\
    GVP-GNN & 0.58 & 0.33 & 0.46 & & 0.80  & 0.61 & 0.81 \\
    \midrule
    OA-GCN    & 0.62 & 0.37 & 0.51 & & 0.84 & 0.65 & 0.85 \\
    OA-GIN    & 0.63 & 0.36 & 0.52 & & \textbf{0.86} & 0.67 & \textbf{0.88} \\
    OA-GAT    & \textbf{0.65} & 0.42 & \textbf{0.53} & & 0.83 & 0.69 & 0.86 \\
    \bottomrule 
    \end{tabular}
    }
\end{minipage}
\vspace{-4.2em}
\end{table}

\begin{table}[!htbp]
\caption{Results of structure biology methods on the TS50.}\label{table:ts50}
\vspace{-0.8em}
\centering
\resizebox{0.97\linewidth}{!}{
\begin{tabular}{@{}lcccccccc@{}}
\toprule
\multicolumn{1}{l|}{Model} & Rosetta & SPIN & ProteinSolver & Wang's & SPIN2 & SBROF & ProDCoNN & St-Trans  \\ \midrule
\multicolumn{1}{l|}{Recv \%} & 30.0 & 30.3 & 30.8 & 33.0 & 33.6 & 39.2 & 40.7 & 42.3 \\ \midrule
\multicolumn{1}{l|}{Model} & DenseCPD & GVP-GNN & OA-GCN & OA-GAT & OA-GIN & No DW & No Int & No Equi  \\ \midrule
\multicolumn{1}{l|}{Recv \%} & 50.7 & 44.9 & 53.8 & \textbf{54.5} & 52.7 & 46.8 & 48.7 & 49.5 \\ \bottomrule
\end{tabular}
}
\vspace{-1.8em}
\end{table}

\textbf{Baselines.} For the CATH 4.2 dataset, we compare our models with models including \textbf{GVP-GNN}, \textbf{Structured Transformer} and \textbf{Structured GNN} \cite{ingraham2019generative}. In the TS50 dataset, baselines includes 3D CNN methods(\textbf{ProDConn} \cite{zhang2020prodconn}, \textbf{DenseCPD} \cite{qi2020densecpd}), sequence models (\textbf{Wang's model} \cite{wang2018computational}, \textbf{SPIN} \cite{li2014direct}, \textbf{SPIN2} \cite{o2018spin2}), \textbf{GNN} method \cite{strokach2020fast}, and classical statistic methods \cite{schaap2001rosetta,cheng2019estimation}.

\textbf{Results.} As shown in Table \ref{table: cath}, our models improve significantly over other methods on the CATH dataset. This result also suggests that the short-chain and single-chain subsets are more challenging, consistent with the result in  \cite{ingraham2019generative}. As shown in Table \ref{table:ts50}, our models also gain superior performance. These results suggest that adopting OA-GNNs is a more effective way of designing valid protein sequences. And the directed weights are the most important module for the CPD task, as the orientational features are essential for the protein to fold to a given structure.

\subsection{Model Quality Assessment}
\label{smqa}

Model quality assessment (MQA) aims to fit a model approximating the numerical metric used to compare the predicted 3D structure with the native \cite{kwon2021assessment}. 
We randomly split the targets in CASP5-CASP10 \cite{moult2014critical} and sample $50$ decoys for each target for generating the training and validation sets and use the CASP11 stage 2 proteins as the test set to ensure no similar structures are involved. There are $508$ proteins for training, $56$ for validation, and $85$ targets with $150$ decoys for the test. 

\textbf{Metrics.} We regress the global GDT\_TS score \cite{zemla2003lga} of predicted structures. We adopt the correlation between the predicted and real values to measure the performance using three statistical correlations: Pearson's correlation $r$, Spearman's $\rho$, and Kendall's $\tau$. We calculate the correlation for each target and average all the correlations over all the targets. We also calculate the Global correlations by taking the union of all decoy sets \cite{pages2019protein}.

\textbf{Baselines.} \textbf{VoroMQA} \cite{olechnovivc2017voromqa} leverages potential model with protein contact maps, while \textbf{RWplus} \cite{zhang2010novel} relies on physical energy terms. \textbf{SBROD} \cite{karasikov2019smooth} uses hand-crafted features. \textbf{Proq3D} \cite{uziela2017proq3d} employs a FCNs for regression. \textbf{3DCNN} \cite{derevyanko2018deep} and \textbf{Ornate} \cite{pages2019protein} apply 3D CNNs for extracting meaningful features. \textbf{GraphQA} \cite{baldassarre2021graphqa}, \textbf{DimeNet} \cite{klicpera2020directional} and \textbf{GVP-GNN} are the most similar models as ours which adopted GNN-based methods on protein graphs. 

\textbf{Results.} As shown in table \ref{table:casp}, our models outperform other methods on both average and global metrics except for Spearman's correlation $\rho$ where we achieve the second-best, demonstrating our models not only evaluates the quality for the same protein but also works well across different proteins in a more general way. 

\section{Conclusion}
For the first time, we generalize the scalar weights in neural networks to 3D directed vectors for better perceiving geometric features like orientations and angles, which are essential in learning representations from protein 3D structures.
Based on the directed weights $\vec{\rmW}$, we provide a toolbox of operators and a $\vec{\rmW}$-Perceptron Unit encouraging efficient scalar-vector feature interactions.
We also enforce global $\mathrm{SO}(3)$-equivariance on a message passing paradigm using the local orientations naturally defined by the rigid property of each amino acid residue, which can be used for designing more powerful networks.
Our integrated Orientation-Aware Graph Neural Networks achieve comparably better performances on multiple biological tasks than existing state-of-the-art methods. 

\section{Acknowledgements}
This work was supported by the National Key Plan for Scientific Research and Development of China (2023YFC3043300) and China's Village Science and Technology City Key Technology funding.






%
%
%
%
%
%
\bibliographystyle{splncs04}
\bibliography{references}

\newpage
\newpage
\appendix

\section{Proof of Rotation Equivariance}
\label{pr}
A function $\vf$ taking a 3D vector $\vx \in \R^3$ as input is rotation equivariance if applying ant rotation matrix $\mR \in \R^{3\times 3}$ on $\vx$ leads to the same transformations of the output $\vf(\vx)$. Formally, $\vf:\R^3 \to \R^3$ is rotation equivariance by fulfilling:
\begin{equation}
    \mR(\vf(\vx)) = \vf(\mR\vx)
\end{equation}
For notation consistency, we consider each row of the matrix to be an individual 3D vector and use right-hand matrix multiplication here.
When performing equivariant message passing on protein graph for node $i$ with local rotation matrix $\mO_i$, we first transform the vector representations of its neighbors from global reference to its local reference, that is, for a particular neighbor node $j$ with vector feature $\mV_j \in \R^{C \times 3}$, apply our DWNN layers $\vf$, and transform the updated features back to the global reference. For simplicity, we set only one neighbor node $u_j$ for $u_i$.The updated vector representation $\mV_i$ for node $i$ is
\begin{equation}
    \mV_i = [\vf(\mV_j\mO_i^{\mathrm{T}})]\mO_i
\end{equation}
If we apply a global rotation matrix $\mR$ to all vectors in the global frame, the local rotation matrix $\mO_i$ will be transformed to $\mO_i^{'} = \mO_i\mR$, and $\mV_j$ to be $\mV_j^{'} = \mV_j\mR$, now the output $\mV_i^{'}$ is
\begin{align}
    \mV_i^{'} &= [\vf(\mV_j^{'}\mO_i^{'\mathrm{T}})]\mO_i^{'} \\
           &= [\vf(\mV_j\mR\mR^{\mathrm{T}}\mO_i^{\mathrm{T}})]\mO_i\mR \\
           &= [f(\mV_j\mO_i^{\mathrm{T}})]\mO_i\mR
\end{align}
And if, instead, we directly apply the rotation matrix to the original output $\mV_i$, we got
\begin{align}
    \mV_i^{'} &= [\vf(\mV_j\mO_i^{\mathrm{T}})]\mO_i\mR \\
             &= \mV_i\mR
\end{align}
In other words, rotating the input leads to the same transformations to the output, so we can preserve rotation equivariance for vector representations in this way. Note that scalar features remain invariant when applying global rotation, so our message-passing paradigm with global and local coordinate transformations is rotation equivariance.

\section{Additional Results}

\subsection{Iterative Prediction on CPD Task}

We tested the sequential decoding of the prediction from the N-side to the C-side and reported the native sequence recovery as shown in Table \ref{tab:cpd_iter},

\begin{table}[htbp!]
\caption{Iterative prediction results of our models on both CATH (Short, Single, All) and TS50 dataset.}
\label{tab:cpd_iter}
\centering
\begin{tabular}{|c|c|c|c|c|}
\hline
Model       & Short & Single & All  & TS50 \\ \hline
OA-GCN (iter) & 34.1  & 35.7   & 45.3 & 50.9 \\ \hline
OA-GIN (iter) & 34.4  & 37.2   & 45.7 & 52.2 \\ \hline
OA-GAT (iter) & 36.0  & 36.5   & 36.5 & 49.6 \\ \hline
\end{tabular}
\end{table}

The results show that the sequential decoding approach basically degraded the results, but because we introduced the directed weight and geometric operations, our results are still better than other benchmarks. We have added these results in the supplementary material.

\subsection{Protein Function Classification}
\label{pfc}

\textbf{Datasets.} Fold Classification (FOLD) predicts the structure category of a given protein structure. We choose the SCOP v1.75 dataset collected by \cite{murzin1995scop} that organizes structures into 3-layer hierarchical classes. There are $16,712$ proteins covering $7$ major structural types with $1,195$ identified folds. We adopt a reduced dataset from \cite{hou2018deepsf}, and remove homologous sequences between test and training data sets at Family (\textbf{Fam}), Superfamily (\textbf{Sup}) or \textbf{Fold} levels, resulting in three different classification tasks.


The Enzyme-Catalyzed Reaction Classification (REACT) study is a very similar task as it requires the classification of the EC number \cite{webb1992enzyme} of a catalyzed enzyme based on the protein structures. Therefore, we put these two tasks side-by-side to evaluate the performance of different methods. We use the dataset collected by \cite{hermosilla2021intrinsic}, containing $37,428$ proteins from $384$ types of Enzyme-Catalyzed classes, and we split them into training, validation set, test set and ensure no sequence or structure overlaps across different sets. There are $29,215$ structures for training, $2,562$ for validation, and $5,651$ for test.

\textbf{Metrics.}
Following standard benchmarks \cite{diehl2019edge,hermosilla2021intrinsic}, we use the classification accuracy for evaluating the prediction performance.

\textbf{Baselines.} We compare with \textbf{CNN-based} models and \textbf{GNN-based} models which learn the protein annotations using 3D structures from scratch. For a fair comparison, we also do not include LSTM-based or transformer-based methods, as they all pre-train their models using millions of protein sequences and only fine-tune their models on 3D structures \cite{bepler2019learning,alley2019unified,rao2019evaluating,strodthoff2020udsmprot,elnaggar2020prottrans}.

\textbf{Results.}
As shown in Table \ref{tab:frclf}, our models demonstrate comparable prediction performance in both tasks across different levels of the hierarchy. Classifying the fold and enzyme classes based on the protein structure is one of the key problems in structural biology. Our proposed models can be used as new tools to conduct function annotations for new proteins.

\begin{table}[htbp]
\caption{Results on Fold and React classification}
\vspace{0.3em}
\label{tab:frclf}
\resizebox{0.95\linewidth}{!}{
\begin{tabular}{@{}cccccc@{}}
\toprule
            & \multirow{2}{*}{Architecture} & \multicolumn{3}{c}{FOLD}                            & \multirow{2}{*}{REACT} \\ \cmidrule(lr){3-5}
            &                               & Fold            & Sup             & Fam             &                        \\ \midrule
\cite{hou2018deepsf}       & 1D ResNet                        & 17.0\%          & 31.0\%          & 77.0\%          & 70.9\%          \\
\cite{derevyanko2018deep}      & 3D CNN                        & 31.6\%          & 45.4\%          & 92.5\%          & 78.8\%          \\ \midrule
\cite{kipf2016semi}        & GCN                          & 16.8\%          & 21.3\%          & 82.8\%          & 67.3\%          \\
\cite{diehl2019edge}       & GCN                          & 12.9\%          & 16.3\%          & 72.5\%          & 57.9\%          \\
\cite{hermosilla2021intrinsic}       & GCN                        & 45.0\%          & 69.7\% & 98.9\%          & 87.2\%          \\
\cite{baldassarre2021graphqa} & GCN                          & 23.7\%          & 32.5\%          & 84.4\%          & 60.8\%          \\
\cite{gligorijevic2021structure}        & LSTM+GCN                     & 15.3\%          & 20.6\%          & 73.2\%          & 63.3\%          \\
\midrule
Our method        & OA-GCN                        & 31.2\%          & 39.8\% & 84.8\%          & 76.0\%          \\
Our method        & OA-GIN                        & 32.8\% & 38.3\%          & 85.2\% & 76.7\% \\
Our method        & OA-GAT                        & 29.9\%          & 36.6\%          & 79.0\%          & 77.2\%          \\ \bottomrule
\end{tabular}
}
\end{table}

For FOLD and REACT tasks, all GNN-based baselines use 3D structure information:
\begin{itemize}
    \item     \cite{kipf2016semi} and \cite{diehl2019edge} construct the input graph using the protein's contact map and use pair-wise distances of residues as edge features.
    \item     \cite{baldassarre2021graphqa} use the spatial features, including the dihedral angles, surface accessibility, and secondary structure type of each residue's node feature. They also use distances, sequence distances, and bond types for edge features.
    \item     \cite{hermosilla2021intrinsic} consider more fine-grained atom-level spatial information, e.g., covalent radius, van der Waals radius, and atom mass, which leads to better performance.
    \item    \cite{gligorijevic2021structure} refer to different types of contact maps and geometric distances for the input of GNN.
\end{itemize}

Despite not using special training techniques, pre-training tasks, and fine-tuning of hyperparameters, our models demonstrate comparable prediction performance in both tasks across different levels of the hierarchy. The introduction of our directed weight with equivariant message passing is especially a big improvement over the normal GNN. 

\section{Experiment Details}
\label{md}
We represent a protein 3D structure as an attributed graph, with each node and edge attached with scalar and geometric-aware vector features. We implement our DWNN in the equivariant message passing manner, with 3 layer Directed Weight Perceptrons for all tasks.

\subsection{Protein Features}
\label{pf}

In this paper, we use an attributed graph $\mathcal{G}=(\mathcal{V},\mathcal{E})$ to represent the protein structure, where each node corresponds to one particular amino acid in the protein with edges connecting its k-nearest neighbors. Here, we set $k=30$. The node features $\mathcal{V}=\{v_1,..,v_N\}$ and edge features $\mathcal{E}=\{e_{ij}\}_{i\neq j}$ are both multi-channel scalar-vector tuples with scalar features like distances and dihedral angles and vector features like unit vectors representing particular orientations.

If available, a node $v_i$ represents the $i$-th residue in the protein with scalar and vector features describing its geometric and chemical properties. Therefore, a node in this graph may have multi-channel scalar-vector tuples $(\vs_i,\mV_i),\vs_i \in \mathbb{R}^6$ or $\mathbb{R}^{26},\mV_i \in \mathbb{R}^{3\times3}$ as its initial features.
\begin{itemize}
    \item \textbf{Scalar Feature.} The $\{ \sin,\cos \} \circ \{\psi,\omega,\phi  \}$. Here $\{\psi,\omega,\phi  \}$ are dihedral angles computed from its four backbone atom positions, $C\alpha_{i-1},N_i,C\alpha_i,N_{i+1}$.
    
    \item \textbf{Scalar Feature.} A one-hot representation of residue if the identity is available.
    
    \item \textbf{Vector Feature.} The unit vectors in the directions of $C\alpha_{i+1}-C\alpha_{i}$ and $C\alpha_{i-1}-C\alpha_{i}$.
    
    \item \textbf{Vector Feature.} The unit vector in the direction of $C\beta_{i}-C\alpha_{i}$ corresponds to the side-chain directions.

\end{itemize}

The edge $e_{ij}$ connecting the $i$-th residue and the $j$-th residue also has multi-channel scalar-vector tuples as its feature $(\vs_{ij},\mV_{ij}),\vs_{ij}\in \mathbb{R}^{34},\mV_{ij}\in \mathbb{R}^{1\times 3}$
\begin{itemize}
    \item \textbf{Scalar Feature.} The encoding of the distance $\|C\alpha_j-C\alpha_i\|$ using 16 Gaussian radial basis functions with centers spaced between $0$ to $20$ angstroms.
    \item \textbf{Scalar Feature.} The positional encoding of $j-i$ corresponds to the relative position in the protein sequence.
    \item \textbf{Scalar Feature:} The contact signal describes if the two residues contact in the space,$1$ if $\|C\alpha_j-C\alpha_j\|\leq 8$ and $0$ otherwise.
    \item \textbf{Scalar Feature.} The H-bond signal describes if there may be an H-bond between the two nodes calculated by backbone distance.
    \item \textbf{Vector Feature.} The unit vector in the direction of $C\alpha_j-C\alpha_i$.
\end{itemize}

\subsection{Model Details}

\textbf{Synthetic Task.} We uniformly sample points on the sphere, and the ratio of positive and negative samples is 1:1. We consider the length of each vector as one channel scalar feature $\in \R$ and the vector itself as one channel vector feature $\in \R^{1\times 3}$. The VNN and GVP models in this experiment are set to 3 layers. And our DWP is only 1 layer, consisting 1 Directed Linear Module, 1 NonLinear Module, and 1 Directed Interaction Module. We also train a 3-layer MLP by concatenating the scalar and vector feature as a four-channel scalar feature $\in \R^4$ as input. Specifically, VNN, GVP, MLP, and DWP parameter counts are $24$,$28$,$24$, and $24$. We use Binary CrossEntropy Loss for this two-class classification task.  For MLP, the scalar and vector features are concatenated along the channel, meaning each point's feature is a vector in $\R^4$. For GVP and VNN, separate neural network components represent the input scalar and vector.

For all models we trained in the protein-related task, we use $128$ scalar channels and $32$ vector channels for each node's hidden representations, $64$ scalar channels, and 16 vector channels for each edge's hidden representations. In implemented models, there are 4 message passing updations, where each passing layer consists of 3 stacked DWPs. In total, our models have about $2.9$ million parameters.

\textbf{CPD.} We train our model in a BERT-style recovery target \cite{strokach2020fast}, more specifically, we dynamically mask 85\% residues of the whole protein and let the model recover them using structure information from its neighbors using CrossEntropy loss. During testing, we mask all the residues and recover the whole protein sequence in one forward pass, we also try to iteratively predict one residue per forwarding pass.

\textbf{MQA.} Because MQA is a regression task, we train our model using MSE loss. When testing, we compute the different types of correlations between predicted scores and ground truth.

\textbf{RES.} Residue identity requires the model to classify the center residue from a local substructure of the protein. We construct a subgraph of each local substructure according to \Secref{pf}. We retrain all the methods based on our protein graph from scratch to guarantee equitable comparison.


\subsection{Training Details}
For the synthetic task, we train each model for 1000 epochs with a learning rate of $1e-3$ and plot the training loss for evaluation.

 We train our models with learning rate $3e-4$ for CPD and $2e-4$ for MQA, a dropout rate of $10\%$ \cite{srivastava2014dropout}, and Layernorm paradigm. We train all the models on NVIDIA Tesla V100 for $100$ epochs for each task.

Our model uses 4-layer message passing, with each message passing consisting of 3-layer DWP. The total model has about 2.9 million parameters.

\subsection{Baseline Details}
We obtain most of our baseline numerical results from other benchmark papers. However, because some models haven’t been evaluated on the datasets we used, we reimplement them using the same model complexity and training hyperparameters as our models. 

In \textbf{the synthetic dataset}, we download the source code from \cite{deng2021vector} and \cite{jing2021equivariant}, both of which use CrossEntropy loss. VNN, GVP, MLP, and DWP parameters are $24,28,24$, and $24$, respectively.

For the RES task \textbf{Table 1}, we implement GCN, GIN, and GAT models, which are trained and evaluated with the same settings as ours. Following the source code from \cite{townshend2020atom3d,jing2021equivariant}, we also retrain 3D CNN and GVP-GNN. To make a fair comparison, we keep the model complexity (e.g. hidden dimensions, layer numbers) and hyperparameters (epochs, random seed) of these baselines the same as our models.

For the CPD task in \textbf{Table 3}, we report the result of St-Transformer from \cite{ingraham2019generative}, St-GCN, and GVP-GNN from \cite{jing2020learning}. We modify the source code from Structed-Transformer\cite{ingraham2019generative} to get the result of St-GAT and St-GIN by replacing the original attention layer in the encoder and decoder module with Graph Attention Network and Graph Isomorphism Network. The hidden dimensions and layer numbers are kept the same as our models. In \textbf{Table 5}, we adopt most of the results from \cite{jing2020learning} except running the St-Transformer model on the TS50 dataset.

For MQA in \textbf{Table 4}, we merge the results of the same task from \cite{jing2020learning} and Structed-Transformer \cite{ingraham2019generative}. GVP-GNN only reports three of six metrics, so we reimplement GVP-GNN and DimeNet on our benchmark. As in the above model, the hidden dimensions and layer numbers are kept the same as in our models. 

For our ablation models, we concatenate scalar and vector features as input and replace all the directed weights with the regular linear layer with scalar weights to get the No DW model. We obtain the No Int model by removing the Interaction Module lying in the final layer of each perceptron. We also replace local frames with unity matrices to break the equivariance in the message passing part (No equivariance).

In the FOLD and REACT tasks, Kipf et al. and Diehl et al. construct the input graph using the protein's contact map and use pair-wise distances of residues as edge features. Baldassarre et al. use spatial features, including dihedral angles, surface accessibility, and secondary structure type, for each residue's node feature. They also use distances, sequence distances, and bond types for edge features. Hermosilla et al. consider more fine-grained atom-level spatial information, e.g., covalent radius, van der Waals radius, and atom mass, which leads to better performance. Gligorijevi et al. refer to different types of contact maps and geometric distances for the input of GNN. All GNN and CNN baselines make use of 3D information in proteins.

\section{Related Work}

\textbf{Representation learning on protein 3D structure.}
Early approaches rely on hand-crafted features extracted and statistical methods to predict function annotations \cite{schaap2001rosetta,zhang2010novel}. Deep learning has been found to achieve success then. 3D CNNs are first proposed to process protein 3D structures by scanning atom-level features using multiple 3D voxels. One of the representative works \cite{derevyanko2018deep} adopts a 3D CNN-based model for assessing the quality of the predicted structures. 3D CNNs also highlight other tasks such as interface prediction \cite{townshend2019end,amidi2018enzynet}. People also extend them to spherical convolutions \cite{gainza2020deciphering,sverrisson2021fast,hermosilla2021intrinsic}, to the Fourier space \cite{zhemchuzhnikov20216dcnn} and the 3D Voronoi Tessellation space \cite{igashov2021vorocnn}. Graph Convolutional Networks \cite{kipf2016semi} have also been adopted to capture geometric and biochemical interactions between residues \cite{ying2018hierarchical,gao2019graph,fout2017protein}, and have been shown to achieve outstanding performance on function prediction \cite{li2021structure}, protein design \cite{strokach2020fast} and binding prediction \cite{vecchio2021neural}. Recently, transformer-based methods \cite{vaswani2017attention} have a trend to replace conventional methods in other bioinformatics tasks \cite{ingraham2019generative,baek2021accurate,jumper2021highly,cao2021fold2seq}.

\textbf{Equivariant neural networks.}
Equivariance is an important property for generalizing to unseen conditions in geometric learning. \cite{cohen2016group,weiler20183d,kohler2020equivariant,satorras2021n,thomas2018tensor,fuchs2020se,anderson2019cormorant,klicpera2020directional,batzner2021se,eismann2021hierarchical}. Methods such as Tensor Field Network \cite{thomas2018tensor} and Cormorant \cite{anderson2019cormorant}, have been developed to generate irreducible representations for achieving rotation equivariance in 3D. In comparison to the Tensor Filed Network with complex tensor products, the Vector Neuron Network (VNN) achieves rotation equivariance in a much simpler way, which generalizes the values of hidden neurons from scalars to 3D vectors \cite{deng2021vector}. GVP-GNN has also been introduced for learning protein representations by featuring geometric factors as vectors \cite{jing2020learning}. Message passing in GNNs for vector representations using equivariant features have also been explored \cite{schutt2021equivariant,luo2022equivariant}. However, to guarantee rotation equivariance, they can only linearly combine 3D vectors, limiting their geometric representing capacity. Another line of methods condition filters on invariant scalar features for maintaining equivariance \cite{schutt2017schnet,gasteiger2019directional,liu2021spherical}.

\end{document}